\input epsf
\documentstyle[12pt,aasms4]{article}

\received{23 May 1996}
\accepted{10 June 1996}

\begin{document}

\title{Comparisons of Cluster Mass Determinations by 
       X-ray Observations and Gravitational Lensing}

\author{Xiang-Ping Wu}
\affil{Department of Physics, University of Arizona, Tucson, AZ 85721 and \\
       Beijing Astronomical Observatory, Chinese Academy of Sciences,
       Beijing 100080, China}

\and

\author{Li-Zhi Fang}
\affil{Department of Physics, University of Arizona, Tucson, AZ 85721}

\begin{abstract}
Gravitational lensing by clusters of galaxies has been detected 
on scales ranging from $\sim10^{-1}$ Mpc to $\sim10$ Mpc, namely,
arcs/arclets, weak lensing and quasar-cluster associations.
This allows us to derive an overall radius matter distribution of
clusters of galaxies. While the dynamical analysis of the 
X-ray observations has yielded a great number of data for
the virial cluster masses, it becomes possible to statistically 
compare the cluster mass determinations by these two independent 
methods. In this letter we show that as compared with gravitational 
lensing, the dynamical analysis under the assumption of isothermal 
and hydrostatic equilibrium has systematically underestimated the 
cluster masses inside the Abell radius by a factor of $\sim2$ 
with  scatter between $0.7$ and $5$. Because the same correction
factor should be applicable to the gas baryon fraction of clusters of 
galaxies obtained from the X-ray data, it is probably too premature 
to claim a baryon crisis in today's cosmology.
\end{abstract}


\section{Introduction}

A combination of the primordial nucleon abundances predicted by   
the standard Big Bang Nucleosynthesis (BBN) and those inferred  
from astronomical observations has set a tight
constraint on the baryonic matter component of the universe 
(Walker et al. 1991): $0.04<\Omega_bh_{50}^{2}<0.06$. 
This indicates that the baryon fraction, 
$f_b\equiv\Omega_b/\Omega$ with $\Omega$ being the average
mass density of the universe in units of critical density, 
is smaller than $\sim0.06\;h_{50}^{-2}$ in the prevailing cosmological 
model of $\Omega=1$, and a significant fraction of the mass
in the universe should be invisible (non-baryonic matter). 
However, such a standard scenario has been challenged
in recent years by the X-ray observations which detect a 
considerably large amount of the hot X-ray emitting gas
in clusters/groups of galaxies. The resulting gas baryon
fraction is a few time greater than the prediction of BBN, 
provided that the gas is in the state of hydrostatic equilibrium
with the gravitational potentials of clusters/groups of galaxies.
The baryon crisis thus arises if the matter 
in clusters/groups of galaxies is representative of the universe.
In particular, this discrepancy probably
implies that at least one of the basic hypotheses
in our current theories of cosmological study needs to be 
modified or even abandoned (White et al. 1993).

Yet, the above claim should be taken very cautiously without
carefully examining the reliability of the X-ray cluster 
mass determinations. Indeed, the existence of substructures
and the recent detection of the complex two-dimensional temperature 
patterns in clusters of galaxies (e.g. Henry \& Briel 1995;
Markevitch 1996; Henriksen \& Markevitch 1996; Henriksen \& White 1996) 
strongly suggest that 
clusters of galaxies may not be the well-virialized 
dynamical systems as were believed before and the uncertainty
in cluster mass determinations assuming hydrostatic equilibrium
for the X-ray gas may be quite large (Balland \& Blanchard 1996).
Therefore, it is desirable that another independent cluster mass 
estimate is made to test the accuracy of the X-ray cluster mass 
determinations and furthermore, to re-examine  whether there is 
a baryon overdensity in clusters of galaxies.

It has been realized that gravitational lensing 
associated with clusters of galaxies can fulfill the task,
which gives rise to cluster masses 
regardless of the cluster matter components and states.
In several clusters of galaxies where both X-ray data and
image distortions of background galaxies are available,
comparisons of the virial cluster masses derived from 
X-ray observations and the gravitational cluster masses
inferred from the distorted images of distant galaxies have been
made (e.g. Wu 1994; Fahlman et al. 1994; Miralda-Escud\'e
\& Babul 1995; Squires et al. 1995).
Today, gravitational lensing by clusters of galaxies has 
been detected on scales ranging from the inner core to the 
outer radius of ten arcminutes, including giant arcs/arclets, 
weakly distorted images of background galaxies and 
quasar-cluster associations. These lensing observations alone
may allow us to derive an overall radius matter distribution
of clusters. Therefore, it would become possible to
statistically compare the cluster matter distributions given 
by dynamical analysis of the X-ray observations with those by
gravitational lensing. This procedure is essentially different from
the previous work which focused on individual clusters 
with both X-ray and gravitational lensing observations. 
We now select the two sets of data separately
from literature. This letter presents the result of the
comparisons and discusses its significance for cosmological study.
Throughout this letter we adopt
a matter-dominated flat cosmological model of $\Omega=1$ and
a Hubble constant of $H_0=50\;h_{50}$ km s$^{-1}$ Mpc.

\section{X-ray cluster mass determination and gas baryon fraction}

Under the assumption of the standard isothermal $\beta$-model for 
the X-ray surface brightness of cluster of galaxies, 
the total mass in gas within radius $r$ of cluster center is
(Cowie, Henriksen, \& Mushotzky 1987) $
M_{gas}(r)=4\pi n_0r_c^3\mu m_p
           \int_0^{r/r_c}x^2(1+x^2)^{-3\beta/2}dx$, 
while the equations of hydrostatic and dynamical equilibrium 
give  the total virial mass $
M_{vir}(r)=3\beta(kT/\mu m_pG)r^3/(r^2+r_c^2)$,
where $n_0$ and $r_c$ denote, respectively, 
the central number density and core radius
of the gas profile, $T$ is the gas temperature,
$k$, the Boltzmann's constant and $\mu m_p$, the mean particle mass.
The ratio of $M_{gas}$ to $M_{vir}$ provides a conservative estimate of
the cluster baryon fraction $f_b$ since the galaxy contribution  
is not included. Fig.1 shows the measured  gas baryon  
fractions of clusters of galaxies in literatures 
without any corrections, in which  we have only utilized  
the virial masses obtained in the case of isothermality. 
Note that for most of the clusters
the gas baryon fractions at radii of larger than $\sim1$ Mpc 
are computed from the spatially-unresolved measurements
of the gas temperature. This leads to an underestimate 
of gas baryon fraction if temperature
decreases with radius as it is naturally expected 
(e.g. Henriksen \& Mamon 1994). 
The relatively low gas baryon fraction at
the largest radius $r\approx7.1$ -- $10$ Mpc for A2142 in Fig.1 
probably arises from such an oversimple assumption
(Henriksen \& White 1996).  It appears that based on the current data,  
we have not detected any apparent variations of 
the gas baryon faction of clusters of galaxies
with radius. The mean gas baryon fractions are 
$\overline{f}_b\approx14\%$ and $\overline{f}_b\approx18\%$ 
with and without those data at $r=0.5$ Mpc given by 
Edge \& Stewart (1991).  It is thus concluded that the
mean baryon fraction in clusters of galaxies is about 2 -- 4 
times larger than the prediction of BBN, if the hot X-ray gas is 
in hydrostatic, isothermal equilibrium with 
the binding cluster gravitational potentials. 

\placefigure{fig1}

\section{Cluster masses from gravitational lensing}

Arcs/arclets are the strongly/moderately distorted images of 
distant galaxies by foreground clusters of galaxies.  
The projected cluster mass within the position ($r_{arc}$) of 
arc/arclet can be easily obtained if one assumes a spherical  
matter distribution for the lensing cluster and approximates 
the alignment parameter of the background galaxy to zero: 
$m_{lens}(r_{arc})=\pi r_{arc}^2\Sigma_{crit}$, where 
$\Sigma_{crit}\equiv (c^2/4\pi G)(D_s/D_dD_{ds})$ is the critical 
mass density with $D_d$, $D_s$ and $D_{ds}$ being the angular 
diameter distances to the cluster, to the galaxy
and from the cluster to the galaxy, respectively. For the complex
arc/arclet configurations, cluster mass can be estimated by constructing
an asymmetrical lens model [see Fort \& Mellier (1994) for a recent review].
We illustrate in Fig.2 the cluster masses given by modeling of arcs/arclets.
One major uncertainty comes from the unknown redshifts for some arclike 
images, for which we have assumed $z_{arc}=0.8$. 
For a typical arc-cluster at redshift of $\sim0.3$, this leads to
an overestimate of cluster mass by a factor of $1.4$ if the background
galaxy is actually located at $z_{arc}=2$.

Another powerful tool of probing the matter distribution of cluster of
galaxies is to study the weak gravitationally induced distortions in
the images of faint galaxies behind cluster of galaxies.
By analyzing the shear field ($\gamma_T$) around the cluster,
the statistics $\zeta(r)=\int_r^{r_{max}}\langle \gamma_T\rangle
(1-r^2/r_{max}^2)^{-1} d\ln r$ 
measures the mean surface mass density 
in units of $\Sigma_{crit}$ interior to $r$
minus that in the annulus from $r$ to $r_{max}$
(Fahlman et al. 1994). Therefore, a lower bound on the projected 
cluster mass within the radius $r$ can be found through 
$m_{lens}(r)=\pi r^2\zeta(r)\Sigma_{crit}$. Cluster mass reconstructions
have been made for several clusters of galaxies in which the 
statistically significant shear patterns are detected. The
resulting cluster masses are plotted in Fig.2. Again, 
there has been so far no information available about the redshifts of 
background galaxies and a mean value of $\langle z\rangle=1$ -- $3$
has been often assumed in the computations. This brings about 
an uncertainty of cluster mass by a factor of $\sim1.3$ 
for a lensing cluster at redshift of $\sim0.3$.\\

Gravitational magnification can also enhance the 
number density of background sources around a foreground 
cluster of galaxies, which has been recently confirmed by 
discovering the so-called quasar-cluster associations 
on scale of up to $\sim10$ arcminutes (Wu \& Fang 1996 and references
therein).  One can figure out the mean cluster mass which is required
to produce the reported quasar overdensity in terms of
gravitational lensing. It turns out that clusters of galaxies should
contain considerably large gravitational masses extending to
a radius of $\sim10$ Mpc in order to account for the quasar enhancements.
We compute the projected cluster masses over the association areas  
simply by $m_{lens}(r)=\pi r^2\Sigma$ and show the results in
Fig.2, where the mean cluster surface mass density $\Sigma$ have 
been given by Wu \& Fang (1996) 
for the four measurements of quasar-cluster associations.

It is remarkable that the projected cluster masses 
revealed statistically by three different lensing methods 
over two decades in radius from $\sim10^{-1}$ Mpc to $\sim10$ Mpc 
can be well fitted by a power-law: 
$m_{lens}(r)=10^{15.39\pm0.17}(r/{\rm Mpc})^{1.51}\;M_{\odot}$,
where (also hereafter)
the error bar represents the scatter of the best-fit
average value rather than the real uncertainty in 
the measurement which is difficult to estimate.
Because the weak lensing method usually provides a 
low limit to the cluster mass, most of its results 
are smaller than the mean value. The fitting 
without weaking lensing data becomes slightly steeper: 
$m_{lens}(r)=10^{15.56\pm0.11}(r/{\rm Mpc})^{1.63}\;M_{\odot}$.  
While one might argue the reliability of the cluster masses
up to the radius of $r\approx10$ Mpc derived from the 
quasar-cluster associations,  we give the fit by removing
the four results of quasar-cluster associations from Fig.2:
$m_{lens}(r)=10^{15.22\pm0.15}(r/{\rm Mpc})^{1.32}\;M_{\odot}$.

\section{Comparisons}

We also display in Fig.2 the X-ray cluster masses $M_{vir}(r)$
derived from the isothermal $\beta$-model under the 
assumption of hydrostatic equilibrium, including the results for five
compact groups of galaxies.   Recall that
$M_{vir}(r)$ are the dynamical masses used in Fig.1 for 
computations of the gas baryon fractions in clusters of galaxies. 
The discrepancy of $M_{vir}$ and $m_{lens}(r)$ is clearly seen at 
small radius, while the two sets of data seem to 
merge beyond $r\sim1$ Mpc. Therefore, an intuitive speculation
for such a variation is the projection effect. 
We have then tested the conventional $r^{-2}$ profile
and the so-called universal density profile found by
the standard CDM simulations (Navarro, Frenk \& While 1995).
Neither of these profiles can fit both the three-dimensional
masses and the two-dimensional projected ones. 
In fact, the deprojection of 
$m_{lens}(r)$ recovers the corresponding 
three-dimensional masses $M_{lens}(r)$, if one 
assumes a spherical matter distribution as 
we have already adopted in the above sections. 
We list the resulting $M_{lens}(r)$ in Table 1, together
with a least-square fit of a power-law to
the X-ray cluster masses $M_{vir}(r)$ 
and the ratios of $M_{lens}(r)/M_{vir}$ at different cluster radii.

\placefigure{fig2}

\placetable{table-1}

There is a significant disprepancy between the virial and
lensing cluster masses inside the core radius of the 
X-ray gas profile, $r_c\sim0.25$ Mpc, where arcs/arclets
are observed. It seems that 
lensing method using arclike images can  give
rise to the cluster masses of  $\sim5$ times 
larger than the virial masses. This is essentially
comparable to the previous similar studies for individual
clusters (Wu 1994, Miralda-Escud\'e \& Babul 1995). 
Around the radius of $r\sim1$ Mpc, at which the lensing data 
are dominated by the weak lensing observations, the mean
ratio of $M_{lens}/M_{vir}$ is about $2$ -- $3$. However,
considering the large scatters, 
one cannot exclude the possibility that the cluster masses
derived from the two methods are consistent. Yet,  most
of the data based on the weaking lensing method correspond 
to the low limits to cluster masses. The similar 
conclusion can be applicable to the Abell radius of $3$ Mpc,
where one may expect a mean factor of $\sim2$ in $M_{lens}/M_{vir}$.
The mass discrepancy might vanish  at the 
outer radii of clusters ($r>5$ Mpc) if we take out the 
results given by the quasar-cluster associations which
dominate the cluster mass determinations with gravitational
lensing at large cluster radius.

\section{Discussion and conclusions}

It appears that as compared with the gravitational lensing method, 
the dynamical analysis based on the isothermal,  
hydrostatic equilibrium has systematically underestimated the 
cluster masses within the Abell radius ($3$ Mpc)
by a factor of $\sim2$ with scatter 
ranging from 0.7 to 5. The mass discrepancy is rather 
remarkable inside the cluster core of $r_c\sim0.25$ Mpc but 
diminishes along the outgoing radius. 
As an immediate result of this discrepancy, 
the baryon fractions in clusters of galaxies provided by the
X-ray cluster masses should be correspondingly reduced by the 
same factor, which thus opens a possibility to 
solve or  partially remove the recently claimed 
baryon crisis in clusters of galaxies. Meanwhile,
our finding indicates that clusters of galaxies may not be 
regarded as the well-relaxed virialized systems. 

The above conclusions are strongly supported by the 
spatially resolved spectra for some clusters of galaxies 
obtained with {\it ASCA}, {\it GINGA} and {\it ROSAT},
which show the complex temperature patterns over the cluster
faces (see Henry \& Briel 1995; Markevitch 1996;
Henriksen \& Markevitch 1996; Henriksen \& White 1996).
These significant temperature 
variations cannot be described by a simple analytic profile
like a $\beta$-model,  and 
non-isothermality in the hot X-ray emitting gas of clusters
of galaxies is apparently required. So, 
cluster mass determinations using the isothermal hypothesis
for the X-ray gas may lead to large errors.

Gravitational lensing is a robust estimate of gravitational 
mass in a celestial body or system.  
We have found the consistency between the cluster masses 
derived from three different lensing phenomena over scale 
$0.1$ Mpc $<r<10$ Mpc.  The total masses inferred from
lensing can be well represented by a single power-law 
of $\sim r^{1.5\pm0.2}$, indicative of a density profile of 
$\sim r^{-1.5\pm0.2}$. This is steeper than the singular isothermal
matter distribution $\sim r^{-2}$ and lies between the
$\sim r^{-1}$ and $\sim r^{-3}$ universal profile predicted by
the standard CDM simulations (Navarro et al. 1995).

Nonetheless, the mass discrepancy between dynamical analysis
and gravitational lensing at the central regions of clusters
of galaxies could arise from the cooling flow and/or the 
contribution of the nonthermal pressure such  as magnetic field
(Loeb \& Mao 1994; Ensslin et al. 1996). Alternatively,
gravitational lensing may overestimate cluster masses
if the lensing cluster is prolate with the major
axis  along the line-of-sight (Miralda-Escud\'e \&
Babul 1995). Furthermore, it is somewhat hard to
understand the cluster mass extension to a radius of
$\sim10$ Mpc where the quasar-cluster associations are detected.

After all, neither of the present X-ray cluster masses 
and gravitational lensing data  forms a complete sample 
and therefore, it is impossible to
evaluate the statistical significance of the result.
What we would like to emphasize in this letter is that 
the comparisons of the updated cluster mass determinations 
by these two independent methods have revealed the possible 
existence of large mass uncertainties under the scenario of 
the hydrostatic equilibrium. 
So, it is probably too premature to claim a baryon
crisis in today's cosmology.

\acknowledgments

We thank an anonymous referee for helpful suggestions.
WXP was supported by the National Science Foundation
of China and a World-Laboratory fellowship.

\clearpage

\begin{table}

\caption{Comparisons of virial and lensing cluster masses.
\label{table-1}}

\bigskip

\begin{tabular}{ccccc}  
\tableline
\tableline
method & power-law fit & 
\multicolumn{3}{c}{$M_{lens}/M_{vir}$}\\
\tableline
 &  & $r=1$ Mpc & $r=3$ Mpc & $r=5$ Mpc \\
\tableline
virial & $M_{vir}(r)=10^{14.53\pm0.23}(r/{\rm Mpc})^{1.91}\;M_{\odot}$ 
            & & & \\
lensing (1) & $M_{lens}(r)=10^{14.96\pm0.06}(r/{\rm Mpc})^{1.51}\;M_{\odot}$ 
            & $2.69_{-1.31}^{+2.56}$ 
            & $1.73_{-0.84}^{+1.65}$ 
            & $1.41_{-0.68}^{+1.35}$  \\
lensing (2) & $M_{lens}(r)=10^{14.90\pm0.03}(r/{\rm Mpc})^{1.32}\;M_{\odot}$ 
            & $2.34_{-1.05}^{+1.93}$  
            & $1.23_{-0.56}^{+1.00}$ 
            & $0.91_{-0.41}^{+0.74}$\\
lensing (3) & $M_{lens}(r)=10^{15.04\pm0.07}(r/{\rm Mpc})^{1.63}\;M_{\odot}$ 
            & $3.24_{-1.62}^{+3.22}$ 
            & $2.38_{-1.19}^{+2.37}$
            & $2.06_{-1.03}^{+2.05}$\\
\tableline
\end{tabular}

\begin{footnotesize}
\tablenotetext{}{(1)Arclike images, weak lensing and quasar-cluster
                  associations; 
                 (2)Arclike images and weak lensing; 
                 (3)Arclike images and quasar-cluster associations}
\end{footnotesize}

\end{table}

\clearpage

\figcaption{Baryon fractions in clusters  of galaxies
derived from the X-ray observations under the assumption of 
isothermal and hydrostatic equilibrium. 
The dashed lines show the predictions of the standard nucleosynthesis
for the cosmological models of $\Omega=1$ and $\Omega=0.3$.
(see Fig.2 for references)
\label{fig1}}

\figcaption{Comparisons of cluster masses inside radius $R$
derived from  dynamical
analysis and gravitational lensing. Open squares: the projected cluster
masses from arcs/arclets (Wu 1994; Kneib \& Soucail 1995); 
Fancy squares: the projected cluster masses from weak lensing 
technique (Fahlman et al. 1994; Tyson \& Fischer 1995; 
Smail \& Dickinson 1995; Luppino \& Kaiser 1996; Seitz et al. 1996; 
Squires et al. 1996a,b); Diamonds: the projected cluster masses from
quasar-cluster associations (Wu \& Fang 1996); Octagons: the cluster
masses from X-ray observations (Hughes et al. 1989;
Edge \& Stewart 1991; Briel, Henry, \& B\"ohringer 1992; 
Miyaji et al. 1993;  White et al. 1993; Briel \& Henry 1994; 
White et al. 1994; Elbaz, Stewart, \& B\"ohringer 1995; 
Dell'Antoio, Geller, \& Fabricant 1995; White \& Fabian 1995; 
Schindler 1995; Schindler \& Wambsganss 1995; Ikebe et al. 1996;
B\"ohringer et al. 1996; Schindler et al. 1996; Squires et al. 1996a;
Henriksen \& White 1996). The masses of compact groups of 
galaxies from X-ray observations are also illustrated (crosses) 
as a comparison (Henriksen \& Mamon 1994; 
Pildis, Bregman, \& Evrard 1995). 
The solid lines show the best fitting power-laws of
the dynamical and lensing data, and the dashed line is the
three-dimensional cluster masses obtained by the deprojection of  
the two-dimensional lensing masses. 
\label{fig2}}

\epsfbox{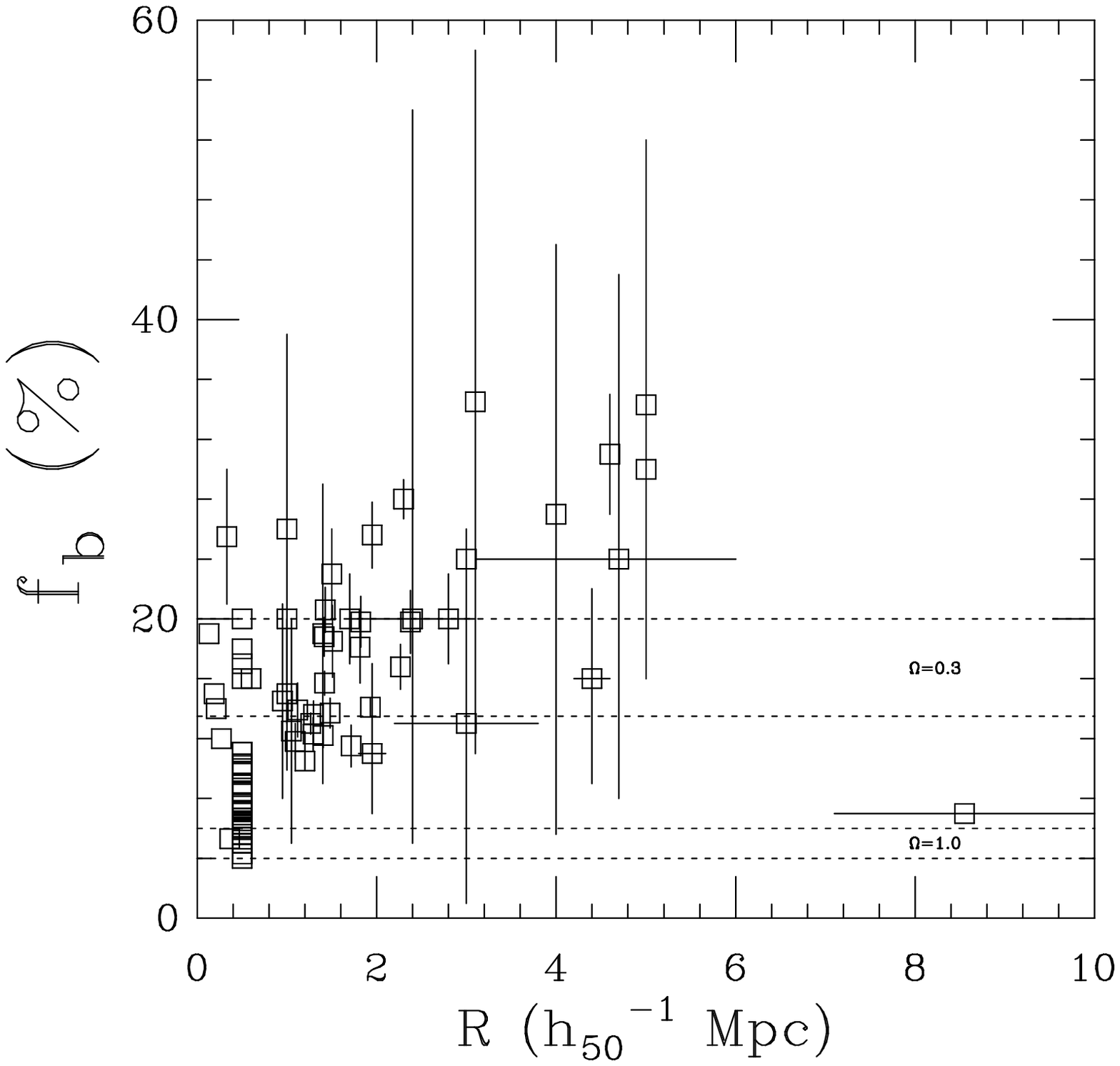}

\epsfbox{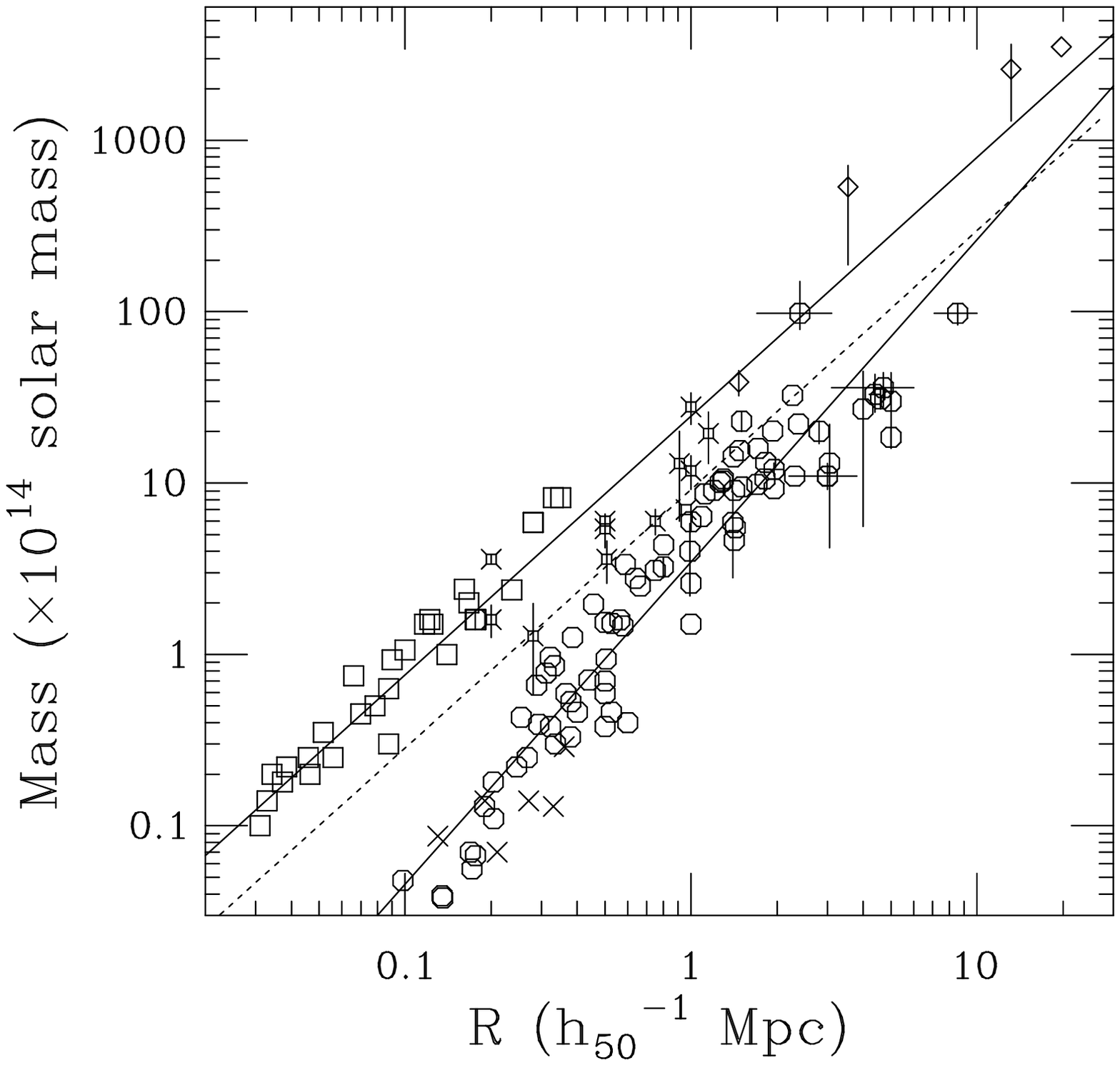}


\clearpage



\begin{references}
\reference{}Balland, C., \& Blanchard, A. 1996, \apj, submitted
\reference{}B\"ohringer, H., Neumann, D. M.,  Schindler, S.,
            \& Kraan-Korteweg, R. C. 1996, \apj, submitted
\reference{}Briel, U. G., \& Henry, J. P., 1994, \nat, 372, 493 
\reference{}Briel, U. G., Henry, J. P., \& B\"ohringer, H. 
            1992, \aap, 259, L31
\reference{}Cowie, L. L., Henriksen, M., \& Mushotzky, R. 
            1987, \apj, 317, 593
\reference{}Dell'Antoio, I. P., Geller, M. J., \& Fabricant, D. G. 
            1995, \aj, 110, 502
\reference{}Edge, A. C., \& Stewart, G. C. 1991, \mnras, 252, 414
\reference{}Elbaz, D., Arnaud, M., \& B\"ohringer, H.
            1995, \aap, 293, 337
\reference{}Ensslin, T. A., Biermann, P. L., Kronberg, P. P.,
            \& Wu, X. P.  1996, \apj, submitted
\reference{}Fahlman, G., Kaiser, N., Squires, G., \& Woods, D.
            1994, \apj, 437, 56
\reference{}Fort, B., \& Mellier, Y. 1994, \aapr, 5, 239
\reference{}Henriksen, M. J., \& Mamon, G. A. 1994, \apj, 421, L63
\reference{}Henriksen, M. J., \& Markevitch, M. L. 1996, \apj, submitted
\reference{}Henriksen, M. J., \& White, III, R. E. 1996, \apj, submitted
\reference{}Henry, J. P., \& Briel, U. G. 1995, \apj, 443, L9
\reference{}Hughes, J. P. 1989, \apj, 337, 21
\reference{}Ikebe, Y., et al. 1996, \nat, 379, 427
\reference{}Kneib, J. P., \& Soucail, G. 1995, 
            Proc of the IAU 173 Symposium on Astrophysical
            Aspects of Gravitational Lensing, ed. 
            C. S. Kochanek, \& J. N. Hewitt (Kluwer 
            Academic Publishers), in press
\reference{}Loeb, A., \& Mao, S. 1994, \apj, 435, L109
\reference{}Luppino, G. A., \& Kaiser, N. 1996, \apj, submitted
\reference{}Markevitch, M. 1996, \apj, in press
\reference{}Miralda-Escud\'e, J., \& Babul, A. 1995, \apj, 449, 18
\reference{}Miyaji, T., et al. 1993, \apj, 419, 66
\reference{}Navarro, J. F., Frenk, C. S., \& White, S. D. M. 1995, 
            \mnras, 275, 720
\reference{}Pildis, R. A., Bregman, J. N., \& Evrard, A. E.
            1995, \apj, 443, 514
\reference{}Seitz, C., Kneib, J.-P., Schneider, P., \& Seitz, S.
            1996, \aap, submitted
\reference{}Schindler, S. 1995, \mnras, submitted
\reference{}Schindler, S., Hattori, M., Neumann, D. M., \&
            B\"ohringer, H. 1996, \aap, submitted
\reference{}Schindler, S., \& Wambsganss, J. 1995, \aap, submitted
\reference{}Smail, I., \& Dickinson, M. 1995, \apj, 455, L99 
\reference{}Squires, G., et al. 1996a, \apj, submitted
\reference{}Squires, G., Kaiser, N., Fahlman, G., Babul, A., 
            \& Woods, D. 1996b, \apj, submitted
\reference{}Tyson, J. A., \& Fischer, P. 1995, \apj, 446, L55
\reference{}Walker, T. P., Steigman, G., Schramm, D. N.,
            Olive, K. A., \& Kang, H.-S. 1991, \apj, 376, 51
\reference{}White, D. A., et al. 1994, \mnras, 269, 589
\reference{}White, D. A., \& Fabian, A. C. 1995, \mnras, 273, 72
\reference{}White, S. D. M., Navarro, J. F., Evrard, A. E.,
            \& Frenk, C. S. 1993, \nat, 366, 429
\reference{}Wu, X. P. 1994, \apj, 436, L115
\reference{}Wu, X. P., \& Fang, L. Z. 1996, \apj, 461, L5
\end{references}
\end{document}